\documentclass[a4paper]{article}
\usepackage{INTERSPEECH2022}
\usepackage{cite}
\usepackage{multirow}

\title{Transferability of Adversarial Attacks on Synthetic Speech Detection}
\name{Jiacheng Deng, Shunyi Chen, Li Dong, Diqun Yan, Rangding Wang}
\address{
  Department of Information Science and Engineering, Ningbo University}

\email{2011082279@nbu.edu.cn, chenshunyi0712@foxmail.com, dongli@nbu.edu.cn, yandiqun@nbu.edu.cn, wangrangding@nbu.edu.cn}

\begin{document}

\maketitle
\begin{abstract}
Synthetic speech detection is one of the most important research problems in audio security. Meanwhile, deep neural networks are vulnerable to adversarial attacks. Therefore, we establish a comprehensive benchmark to evaluate the transferability of adversarial attacks on the synthetic speech detection task.
Specifically, we attempt to investigate: 1) The transferability of adversarial attacks between different features. 2) The influence of varying extraction hyperparameters of features on the transferability of adversarial attacks. 3) The effect of clipping or self-padding operation on the transferability of adversarial attacks.
By performing these analyses, we summarise the weaknesses of synthetic speech detectors and the transferability behaviours of adversarial attacks, which provide insights for future research.
More details can be found at https://gitee.com/djc\_QRICK/Attack-Transferability-On-Synthetic-Detection.
\end{abstract}
\noindent\textbf{Index Terms}: Synthetic speech detection, adversarial attack, adversarial robustness

\section{Introduction}
Synthetic speech detection \cite{yamagishi2021asvspoof} focuses on detecting spoof speech in non-auto speaker verification scenarios. It reflects the scenario in which an attacker has access to the voice data of a targeted victim, e.g., data posted to social media. The victim might be a celebrity, a social media influencer, or an ordinary citizen. The attacker¡¯s incentive might be, e.g., to blackmail the victim or to denigrate his or her reputation in some way by spreading spoken misinformation. Synthetic speech techniques commonly used by attackers include text-to-speech (TTS) \cite{fan2014tts} and speech conversion (VC) \cite{toda2016voice,mohammadi2017overview}. In order to detect this kind of attack, researchers have proposed a variety of countermeasures \cite{sahidullah2015comparison,paul2017spectral,hua2021towards}.

However, the existing methods are only used for unprocessed synthetic speech, and the exploration of the impact of adversarial attacks on synthetic speech identification is somewhat inadequate. Adversarial attacks are algorithms that add imperceptible perturbations to the input signal of the machine learning model to produce incorrect output labels. The perturbed input signal is called a adversarial example. Therefore, adversarial attacks pose a security threat to the deep learning model of synthetic speech detection, speaker recognition, and other tasks.

Considering that there are few works on transferability in synthetic speech detection, we review similar works in sound classification. \cite{du2020sirenattack}\cite{subramanian2020study} studies the transferability of adversarial examples generated based on the Mel-frequency cepstrum coefficient (MFCC) between different deep models. Esmaeilpour \emph{et al.} \cite{esmaeilpour2019robust} proved the transferability of adversarial attacks between deep learning models and Support Vector Machines (SVM). Subramanian \emph{et al.} \cite{subramanian2019robustness} studied the robustness of adversarial samples under different input transformations, such as MP3 compression, resampling, and white noise. Subramanian \emph{et al.} \cite{subramanian2019robustness} implemented the transferability of adversarial attack from spectrogram to raw waveform. Their research does not delve into three problems:
1) Transferability of adversarial attack between various features; 2) Transferability of adversarial attack between the same feature with different extraction hyperparameters; 3) Transferability of adversarial attacks on clipping and self-padding operations. Taking the DF track of ASVspoof2021\cite{yamagishi2021asvspoof} as an example, the classifier usually uses multiple features, and it is possible to extract features based on the raw waveform of different lengths. Furthermore, although some features such as MFCC, LFCC, CQCC, spectrogram, etc., are commonly used, different hyperparameters for feature extraction are used in different models.

Given the lack of research on transferability in synthetic speech detection, it is necessary to deepen an understanding of the properties of defending and attacking using the domain knowledge of the field. In this paper, we conduct extensive transferability experiments across different input features and different lengths of the raw waveform.

The main contributions of our work are summarized as follows:
\begin{enumerate}
  \item We evaluate the transfer behavior of adversarial attacks between MFCCs, LFCCs, spectrograms, and waveforms. The results reveal the vulnerability of the 2D feature-based models and the strong defense capability of the 1D feature-based models against adversarial attacks.
  \item We verify that employing different feature extraction methods does not defend against adversarial attacks.
  \item We investigate the effect of clipping and self-padding operations on the transferability of adversarial attacks. The results indicate that the clipping operation reduces the threat of adversarial attacks, while the self-padding operation does the opposite.
\end{enumerate}

\section{METHODOLOGY}
This section describes the experimental setting, including the adversarial attacks employed, the threatened models, the datasets used, and the evaluation metric.
\subsection{Adversarial attacks}
In this paper, all attacks used are implemented under white-box settings. The attacker can obtain all the information about the model.
We ignore the weak baseline attack method FGSM \cite{goodfellow2014explaining} and adopt the I-FGSM \cite{madry2017towards} and the more robust attack I-FGSM$_{ens}$ \cite{dong2018boosting}. The attacks used are as follows:

\textbf{I-FGSM} is the fast gradient-based iterative attack method for generating adversarial examples, and it has a high attack success rate.
Assume $J(x,y)$ as cross-entopy cost function of the model, given wave $X$ and targeted class $y$. The I-FGSM can be formulated as follow:
\begin{equation}
    \begin{aligned}
    x^{adv}_0&=x,\\
    x^{adv}_{N+1}&=\rm{Clip}_{x,\epsilon}\{x^{adv}_{N}-\alpha \texttt{sign}(\nabla_x J(x^{adv}_N,y))\}
    \end{aligned}
\end{equation}
where $x^{adv}$ is an adversarial example, $x$ is source audio, $\alpha$ is the step size. $\rm{Clip}_{x,\epsilon}\{x'\}$ is function that makes $x'$ in $L_\infty$ $\epsilon$-neighbourhood of $x$.

$\textbf{\rm{\textbf{I-FGSM}}}_{ens}$ is an ensemble attack based on I-FGSM. \cite{wang2021enhancing,zhang2021generating,wang2021boosting} show that ensemble attacks can greatly improve the transferability of adversarial attacks. Following the setting of \cite{dong2018boosting}, assuming there are $K$ models, the loss function of I-FGSM$_{ens}$ is defined as follows:
\begin{equation}
J(x,y)=\sum^K_{k=1}\nabla_{x}J(x,y)
\end{equation}

For I-FGSM, set $\epsilon=0.08$, $\alpha=0.001$ and the number of iterations to 40. For I-FGSM$_{ens}$, set $\epsilon=0.1$, $\alpha=0.002$ and the number of iterations to 60.

\subsection{Input Features}
\label{sec:input-feature}
\begin{figure}[h]
  \centering
  \includegraphics[width=\linewidth]{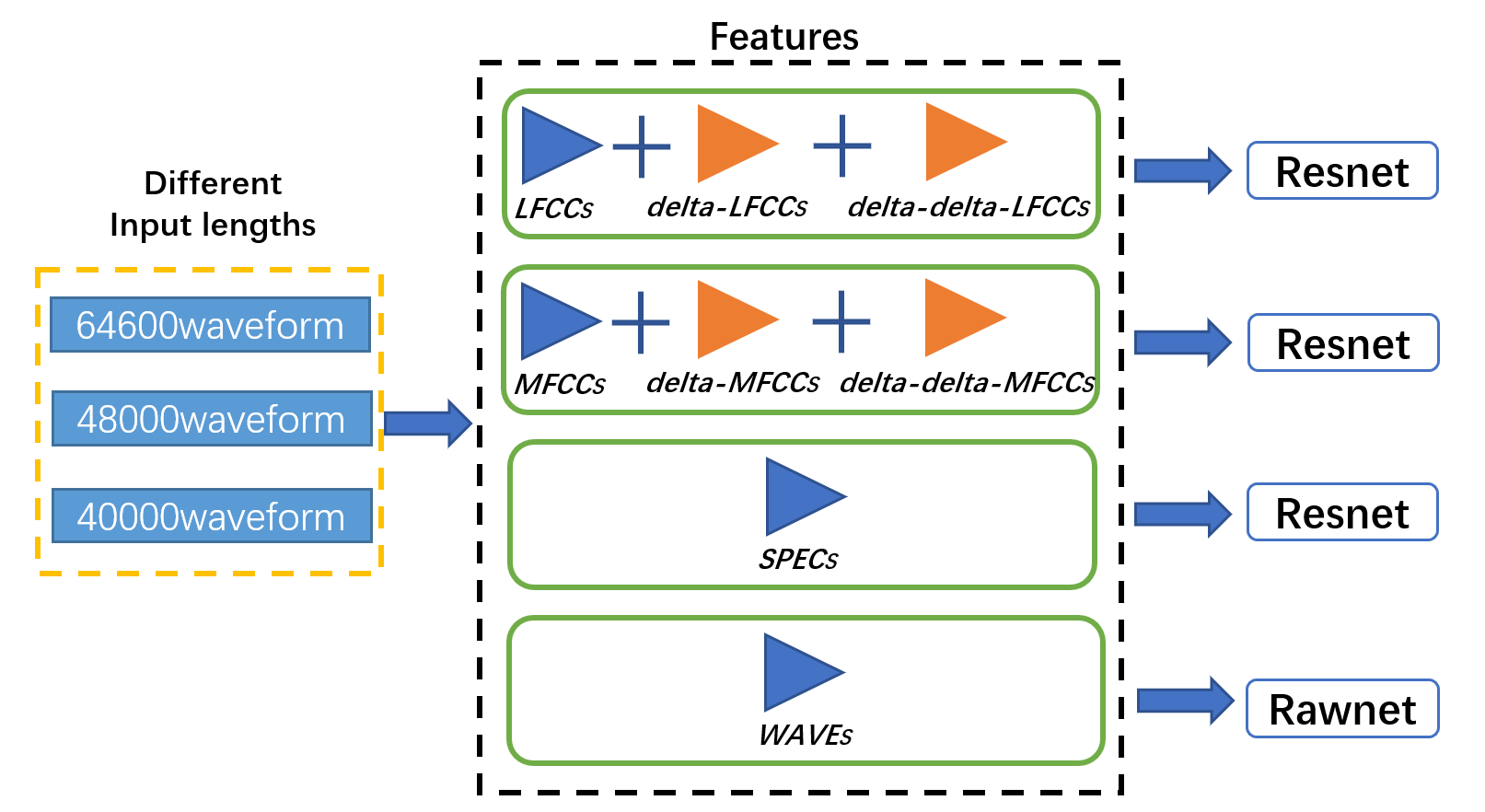}
  \caption{Input, features and models.}
  \label{fig:features}
\end{figure}

As shown in Figure.\ref{fig:features}, we choose several features widely used in synthetic speech classification, including LFCCs \cite{luo2021capsule,sahidullah2015comparison}, MFCCs \cite{tiwari2010mfcc,kumar2011delta,hossan2010novel}, Spectrogram (SPEC) \cite{sahidullah2015comparison}, raw waveform (WAVEs)\cite{todisco2019asvspoof,jung2019rawnet}, delta, and delta-delta \cite{hossan2010novel}. Considering multiple extraction methods for two-dimensional(2D) features, we extract each feature three times with different hyperparameters.

\textbf{LFCCs:} Linear frequency cepstral coefficients are a popular feature used for speech classification. The LFCC process has two steps: 1)Using linear filters and logarithmic compression power spectrum; 2)Perform DCT to produce the cepstral coefficients. In the experiment, we adjust the retaining LFCC coefficients to produce three features, $\rm{LFCC}_{60}$, $\rm{LFCC}_{70}$, and $\rm{LFCC}_{80}$.

\textbf{MFCCs:} In the process of Mel-frequency cepstral coefficients, Mel scale filters are used to replace the linear filters in LFCC, having denser spacing in the low-frequency region. MFCC is sensitive to the change of low frequency. In the experiment, like LFCC, we adjust retaining MFCC coefficients to produce three features, $\rm{MFCC}_{30}$, $\rm{MFCC}_{40}$, and $\rm{MFCC}_{80}$.

\textbf{SPECs:} Spectrogram contains rich information related to speech or speaker\cite{rabiner1993fundamentals}. We use the log-spectrogram calculated directly from the speech frame as the input feature. The specific parameter settings are as follows: win\_length=1024, hop\_length=512, Hann window \cite{zierhofer2007data}, n\_fft=1024, 2048 or 3072. We abbreviate the SPECs from different n\_fft as $\rm{SPEC}_{1024}$, $\rm{SPEC}_{2048}$, and $\rm{SPEC}_{3072}$.

\textbf{WAVEs:} Input acoustic sampling values directly into the model. It is essential to explore the transferability of adversarial attacks between one-dimensional(1D) and 2D features.

\textbf{delta and delta-delta:} They can record some dynamic information \cite{soong1988use}, such as the change of MFCC and LFCC over time. Therefore, when MFCCs or LFCCs are used as features, we input delta-MFCCs/LFCCs and delta-delta-MFCCs/LFCCs at the same time.

\subsection{Dataset}
We use the ASVspoof2021 dataset \cite{wang2020asvspoof} introduced for the speech deepfake challenge. The training dataset contains 2580 bona fide speech and 22800 spoofed speech. The test dataset contains 24844 speeches. The adversarial examples we generate are based on spoof speech samples which is the meaningful scenario that adversarial synthetic speech is recognized as bona fide speech.

\subsection{Model}

\begin{table}[h]
\caption{Average accuracy of models based on different features}
\label{tab:model-accuracy}
\centering
\begin{tabular}{c|cc}
\hline
Feature & Train(\%) & Test(\%) \\ \hline
LFCCs    & 99.8      & 99.8     \\
MFCCs    & 99.7      & 99.6     \\
SPECs    & 99.9      & 99.9     \\
WAVEs    & 99.4      & 99.0    \\ \hline
\end{tabular}
\end{table}

We input two-dimensional features into the Resnet \cite{he2016deep} model and one-dimensional features into the Rawnet \cite{jung2019rawnet} model. Resnet is a typical 2D CNN model, which has superior performance in the classification of two-dimensional features such as spectrogram. Rawnet is direct modelling of raw waveforms using deep neural networks, which has achieved state-of-the-art in speaker classification and synthetic speech identification. As shown in Figure.\ref{fig:features}, we feed the models with three different input dimensions including 1$\times$40000, 1$\times$48000, and 1$\times$64600 to simulate clipping and self-padding. The models under different input dimensions are abbreviated to $\rm{res}_{646}$, $\rm{res}_{48}$, $\rm{res}_{4}$, $\rm{raw}_{646}$, $\rm{raw}_{48}$, and $\rm{raw}_{4}$.

Table \ref{tab:model-accuracy} provides average accuracy of models based on different features. Taking MFCCs as an example, its result is the average accuracy of $\rm{MFCC}_{30}+res_{646}$, $\rm{MFCC}_{30}+res_{48}$, $\rm{MFCC}_{30}+res_{4}$, $\rm{MFCC}_{40}+res_{646}$, $\rm{MFCC}_{40}+res_{48}$, $\rm{MFCC}_{40}+res_{4}$, $\rm{MFCC}_{80}+res_{646}$, $\rm{MFCC}_{80}+res_{48}$, and $\rm{MFCC}_{80}+res_{4}$.

\subsection{Metrics}

\begin{table*}[t]
\caption{The TSR of adversarial attacks between different features}
\label{tab:feature}
\centering
\begin{tabular}{c|c|ccccc}
\hline
Attack &
  Model &
  \begin{tabular}[c]{@{}c@{}}SNR\\ (dB)\end{tabular} &
  \begin{tabular}[c]{@{}c@{}}$\rm{LFCC}_{70}$\\ +$\rm{res}_{646}$\end{tabular} &
  \begin{tabular}[c]{@{}c@{}}$\rm{MFCC}_{40}$\\ +$\rm{res}_{646}$\end{tabular} &
  \begin{tabular}[c]{@{}c@{}}$\rm{SPEC}_{2048}$\\ +$\rm{res}_{646}$\end{tabular} &
  \begin{tabular}[c]{@{}c@{}}WAVE+\\$\rm{raw}_{646}$\end{tabular}  \\ \hline
\multirow{4}{*}{\begin{tabular}[c]{@{}c@{}}I-\\ FGSM\end{tabular}}   & $\rm{LFCC}_{70}+res_{646}$ & 39.4  & 1.0 & 1.0 & 0.98  & 0.0   \\
                                                                     & $\rm{MFCC}_{40}+res_{646}$ & 38.7  & 1.0  & 1.0 & 0.99  & 0.0  \\
                                                                     &$\rm{SPEC}_{2048}+res_{646}$& 38.2  & 0.96 & 0.97 & 1.0  & 0.0  \\
                                                                     & WAVE+$\rm{raw}_{646}$      & 39.5  & 0.64 & 0.39 & 0.99 & 1.0   \\ \hline
\begin{tabular}[c]{@{}c@{}}I-\\ $\rm{FGSM}_{ens}$\end{tabular}                  &$\rm{res}^*_{646}+\rm{raw}_{646}$                  &  34.3   &  1.0    & 1.0     & 1.0     &  0.61    \\ \hline
\end{tabular}
\end{table*}

\begin{table*}[t]
\caption{The TSR of adversarial attacks between different features with different hyperparameters}
\label{tab:feature2}
\centering
\begin{tabular}{c|c|ccccclll}
\hline
Attack &
  Model &
  \begin{tabular}[c]{@{}c@{}}$\rm{LFCC}_{60}$\\ +$\rm{res}_{646}$\end{tabular} &
  \begin{tabular}[c]{@{}c@{}}$\rm{MFCC}_{30}$\\ +$\rm{res}_{646}$\end{tabular} &
  \begin{tabular}[c]{@{}c@{}}$\rm{SPEC}_{1024}$\\ +$\rm{res}_{646}$\end{tabular} &
  \begin{tabular}[c]{@{}l@{}}$\rm{LFCC}_{80}$\\ +$\rm{res}_{646}$\end{tabular} &
  \begin{tabular}[c]{@{}l@{}}$\rm{MFCC}_{80}$\\ +$\rm{res}_{646}$\end{tabular} &
  \begin{tabular}[c]{@{}l@{}}$\rm{SPEC}_{3072}$\\ +$\rm{res}_{646}$\end{tabular} \\ \hline
\multirow{4}{*}{\begin{tabular}[c]{@{}c@{}}I-\\ FGSM\end{tabular}}   & $\rm{LFCC}_{70}+res_{646}$ & 1.0  & 0.87 & 0.83 & 1.0  & 1.0  & 0.98   \\
                                                                     & $\rm{MFCC}_{40}+res_{646}$ & 0.99  & 0.99  & 0.99 & 0.99  & 1.0 & 0.99   \\
                                                                     &$\rm{SPEC}_{2048}+res_{646}$& 0.98  & 0.74 & 0.94 & 0.98  & 0.96 & 1.0  \\
                                                                     & WAVE+$\rm{raw}_{646}$           & 0.68  &0.22  & 0.76 & 0.59 & 0.35  & 0.99   \\ \hline
\begin{tabular}[c]{@{}c@{}}I-\\ $\rm{FGSM}_{ens}$\end{tabular}                  &$\rm{res}^*_{646}+\rm{raw}_{646}$                  &  0.99  &  0.96  &  0.93  &  1.0 &  0.99   & 1.0   \\ \hline
\end{tabular}
\end{table*}

The experimental evaluation criteria include attack success rate (ASR), the quality of adversarial examples, and transfer success rate (TSR).

\textbf{ASR:} It is worth stating that all adversarial examples are generated based on spoof synthetic speech. Considering that there may be a small amount of misclassification in the model itself, the attack success rate for model M is calculated as follows:
\begin{equation}
\rm{ASR}_{M}=\frac{\rm{NUM}(\hat{X}^M_{X^{All}- X^{err}_M})}{\rm{NUM}(X^{All} - X^{err}_M)},
\end{equation}
where NUM is a count function, $\rm{X}^{All}$ is the dataset, $\rm{X}^{err}_M$ is the sample set misclassified by model M, and $\rm{\hat{X}}^a_b$ is the adversarial examples generated based on data set $b$ and model $a$.

\textbf{TSR:} Assuming that there are model M and model N, the transferability of adversarial attacks from M to N can be formulated as follows:
\begin{equation}
\begin{aligned}
    \rm{S}&=\rm{\hat{X}}^{M}_{X^{All}-X^{err}_{M}},\\
    \rm{TSR}^{N}_{M}&=\frac{\rm{NUM}(\hat{X}^{N}_{S})}{\rm{NUM}(S - X^{err}_{N})},
\end{aligned}
\end{equation}

\textbf{SNR:} Signal-to-noise ratio \cite{johnson2006signal} is an objective metric to evaluate the quality of adversarial examples \cite{du2020sirenattack,kereliuk2015deep,abdoli2019universal}, and it is defined as:
\begin{equation}
  \rm{SNR}_{dB}(\emph{r},\emph{x})=10log_{10}\frac{\sqrt{\sum^N_{n=1}\emph{x}^2_n}}{\sqrt{\sum^N_{n=1}\emph{r}^2_n}}
\end{equation}
where $r$ is noise audio and $x$ is clean audio.

\section{Disscussion}

\subsection{ASR and SNR of adversarial attacks}
\begin{figure}[h]
  \centering
  \includegraphics[width=0.8\linewidth]{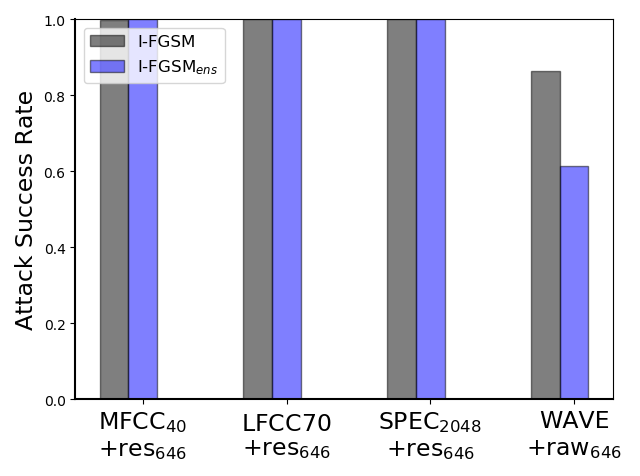}
  \caption{Attack Success Rate.}
  \label{fig:ASR}
\end{figure}

Table \ref{tab:feature} and Fig.\ref{fig:ASR} show the SNR and ASR respectively. Generally, the SNR exceeding 20dB is not perceptible to the human ear, and the SNR generated in Table \ref{tab:feature} is between 39.4dB and 34.3dB, which shows that the adversarial examples generated in this paper are of high quality. Fig.\ref{fig:ASR} shows the success rates of different attacks on the 64600 series of models. The average attack success rate for I-FGSM/I-FGSM$_{ens}$ is 92.5\%/90.5\%, which indicates that adversarial attacks threaten models with different features.

\subsection{TSR between different features}

\begin{table*}[t]
\caption{Effect of clipping operation on the TSR of adversarial attacks.}
\label{tab:clipping}
\centering
\begin{tabular}{c|c|cccccccc}
\hline
Attack &
  Model &
  \begin{tabular}[c]{@{}c@{}}$\rm{LFCC}_{70}$\\ +$\rm{res}_{48}$\end{tabular} &
  \begin{tabular}[c]{@{}c@{}}$\rm{MFCC}_{40}$\\ +$\rm{res}_{48}$
  \end{tabular} &
  \begin{tabular}[c]{@{}c@{}}$\rm{SPEC}_{2048}$\\ +$\rm{res}_{48}$\end{tabular} &
  \begin{tabular}[c]{@{}c@{}}WAVE\\ +$\rm{raw}_{48}$\end{tabular}   &
  \begin{tabular}[c]{@{}c@{}}$\rm{LFCC}_{70}$\\ +$\rm{res}_{4}$\end{tabular}   &
  \begin{tabular}[c]{@{}c@{}}$\rm{MFCC}_{40}$\\ +$\rm{res}_{4}$\end{tabular}  &
  \begin{tabular}[c]{@{}c@{}}$\rm{SPEC}_{2048}$\\ +$\rm{res}_{4}$\end{tabular} &
  \begin{tabular}[c]{@{}c@{}}WAVE\\ +$\rm{raw}_{4}$\end{tabular}   \\ \hline
\multirow{4}{*}{\begin{tabular}[c]{@{}c@{}}I-\\ FGSM\end{tabular}}   & $\rm{LFCC}_{70}+res_{646}$ & 1.0   & 0.99  &  0.97 & 0.0 & 1.0 & 0.98&0.81 & 0.0 \\
                                                                     & $\rm{MFCC}_{40}+res_{646}$ & 0.93  & 1.0   & 0.95 & 0.0 & 0.98& 1.0 &0.89 & 0.0 \\
                                                                     & $\rm{SPEC}_{2048}+res_{646}$ & 0.95& 0.96  & 1.0 & 0.0 & 0.73& 0.86& 0.95 & 0.0\\
                                                                     & WAVE+$\rm{raw}_{646}$ & 0.23  & 0.39 & 0.97 & 0.01 & 0.07 & 0.02 &0.82 &0.01\\ \hline
\begin{tabular}[c]{@{}c@{}}I-\\ $\rm{FGSM}_{ens}$\end{tabular}                  &$\rm{res}^*_{646}+\rm{raw}_{646}$& 1.0 & 0.99 & 1.0  & 0.0& 0.99 & 0.98 & 0.69 & 0.0 \\ \hline
\end{tabular}
\end{table*}

\begin{table*}[t]
\caption{Effect of self-padding operation on the TSR of adversarial attacks.}
\label{tab:self-padding}
\centering
\begin{tabular}{c|c|cccccccc}
\hline
Attack &
  Model &
  \begin{tabular}[c]{@{}c@{}}$\rm{LFCC}_{70}$\\ +$\rm{res}_{48}$\end{tabular} &
  \begin{tabular}[c]{@{}c@{}}$\rm{MFCC}_{40}$\\ +$\rm{res}_{48}$
  \end{tabular} &
  \begin{tabular}[c]{@{}c@{}}$\rm{SPEC}_{2048}$\\ +$\rm{res}_{48}$\end{tabular} &
  \begin{tabular}[c]{@{}c@{}}WAVE\\ +$\rm{raw}_{48}$\end{tabular}   &
  \begin{tabular}[c]{@{}c@{}}$\rm{LFCC}_{70}$\\ +$\rm{res}_{646}$\end{tabular}   &
  \begin{tabular}[c]{@{}c@{}}$\rm{MFCC}_{40}$\\ +$\rm{res}_{646}$\end{tabular}  &
  \begin{tabular}[c]{@{}c@{}}$\rm{SPEC}_{2048}$\\ +$\rm{res}_{646}$\end{tabular} &
  \begin{tabular}[c]{@{}c@{}}WAVE\\ +$\rm{raw}_{646}$\end{tabular}   \\ \hline
\multirow{4}{*}{\begin{tabular}[c]{@{}c@{}}I-\\ FGSM\end{tabular}}   & $\rm{LFCC}_{70}+res_{4}$ & 1.0 &1.0   & 0.97 & 0.0   & 1.0 & 1.0 &0.98 &0.0  \\
                                                                     & $\rm{MFCC}_{40}+res_{4}$ & 0.96  & 1.0 & 0.97   &  0.0&1.0  &1.0  &0.98 & 0.0 \\
                                                                     & $\rm{SPEC}_{2048}+res_{4}$ & 1.0  &0.98 & 0.99     & 0.0 & 0.99 & 0.98 & 1.0& 0.01\\
                                                                     & WAVE+$\rm{raw}_{4}$ & 0.61  & 0.73 & 0.98 & 0.0 & 0.45 & 0.71 &0.99 &0.01\\ \hline
\begin{tabular}[c]{@{}c@{}}I-\\ $\rm{FGSM}_{ens}$\end{tabular}                  &$\rm{res}^*_{4}+\rm{raw}_{4}$& 1.0      & 1.0 & 1.0  & 0.0   & 1.0 & 1.0 & 0.99 & 0.0 \\ \hline
\end{tabular}
\end{table*}
Table \ref{tab:feature} and Table \ref{tab:feature2} show the TSR of adversarial attacks between MFCCs, LFCCs, SPECs, and WAVEs.
We observe that average TSR from MFCCs to LFCCs is 99.3\%, the average TSR from LFCCs to MFCCs is 95.6\%, the average TSR from MFCCs/LFCCs to SPECs achieves 99\%/93\% , and the average TSR from SPECs to MFCCs/LFCCs is 97.3\%/89\%.
These results strongly indicate that adversarial attacks are highly transferable between 2D features. Furthermore, the lowest TSR from different features to SPECs is 76\%, and the average TSR is 95.3\%, which indicates that spectrogram features are more vulnerable than other features.
Taking the DF track of ASVspoof2021 as an example, 2D features are widely used and are the primary input features of various classifiers. Therefore, our experiments demonstrate the vulnerability of models based on 2D features to the threat of 2D-features-based adversarial attacks.

Meanwhile, we find that the transferability of adversarial attacks from 2D-feature-based models to 1D-feature-based models is almost 0\%, whereas adversarial attacks based on 1D features still threaten 2D-feature-based models. The above results guide both attackers and defenders in the synthetic speech detection task. For the defender, the classifier, the use of 1D features can well defend against adversarial attacks based on 2D features. For the attacker, the adversarial attack based on 1D features can threaten the classifier with the 1D feature and the classifier based on the 2D features.

\subsection{TSR between different features with different extraction hyperparameters}
This experiment is devoted to investigating whether changes in the hyperparameters of feature extraction affect transferability. We follow Section \ref{sec:input-feature} to adjust the extraction hyper-parameters of the features. In Table \ref{tab:feature2}, the average TSR of adversarial attacks drops by 1.2\% compared to Table \ref{tab:feature}.
Meanwhile, we observe that the average TSR from MFCCs/LFCCs/SPECs/WAVEs to 2D-features is 99.1\%/94.6\%/93.3\%/59.8\%, and the average TSR from LFCCs/SPECs/WAVEs to MFCCs is 93.5\%/85\%/28.5\%.
The results show: 1) Adversarial attacks are transferable between features based on different extracted parameters; 2) Changing the feature extraction parameters has a limited effect on the TSR; 3) Adversarial attacks based on MFCCs have superior transferability for 2D-feature-based models.

\subsection{TSR on clipping and self-padding operation}
Different classifiers usually take raw waveforms of different lengths as input and extract features. Therefore, we evaluate the correlation between the length of the input raw waveform and the TSR of the adversarial attack. Table \ref{tab:clipping} clips the adversarial examples generated by the 64600-series model into 48000 and 40000 lengths. The clipping operation is to select the first 40000/48000 sampling points of each adversarial example. Comparing the TSR in Tables \ref{tab:feature} and \ref{tab:clipping}, the 646-series-based adversarial attack drops 7.25\% on the 48-series model and 9\% on the 4-series model. Specifically, adversarial attacks based on WAVE+res$_{646}$ can hardly threaten LFCCs/MFCCs+res$_4$. This result shows that adversarial examples are susceptible to clipping operations, especially those generated based on the raw waveform.

In Table \ref{tab:self-padding}, we self-padded the adversarial examples generated based on the 4-series model into 48000 and 64600 lengths. The self-padding operation is to concatenate each adversarial example with its own first 8000/24600 sampling points. Comparing the TSR in Tables \ref{tab:feature} and \ref{tab:self-padding}, the 4-series-based adversarial attack increases by 3.9\% on the 48-series model and 3.3\% on the 646-series model. In particular, the TSR of adversarial attacks from WAVE+raw$_{646}$ to LFCCs/MFCCs+res$_{48}$/res$_{4}$ was significantly improved. This result shows that the self-padding operation does not affect the TSR of adversarial attacks and increases the threat of adversarial attacks based on 1D features.

\subsection{TSR of ensemble attack}
Ensemble adversarial attack is an effective method to improve the threat of attack. In Table \ref{tab:feature} and Table \ref{tab:feature2}, the average TSR of I-FGSM$_{ens}$ is 86.1\%, and the average TSR of I-FGSM is 79.8\%, which shows that the ensemble adversarial attack can effectively improve the transferability.
However, Tables \ref{tab:clipping} and \ref{tab:self-padding} demonstrate that ensemble-based adversarial attacks cannot threaten models based on 1D features of different lengths, which illustrates the strong robustness of WAVEs feature-based models.

\section{Conclusions}
This paper scratches the transferability of adversarial attacks in synthetic speech detection under different settings. These settings simulate real detection scenarios: 1) Adopt various features as model input, such as MFCCs, LFCCs, spectrograms, and waveforms; 2) Use different feature extraction parameters, such as filter numbers; 3)Take raw waveforms of different lengths as input and extract features. Based on the evaluation results, we draw some critical findings of adversarial attacks transfer behavior, which may be helpful for future research.

 In the future,  our work will focus on improving the transferability of adversarial attacks between the different lengths of 1D features from the attacker's perspective.

\bibliographystyle{IEEEtran}


\end{document}